\documentclass{aastex63}

\usepackage{color}
\UseRawInputEncoding
\received{}%{June 1, 2019}
\revised{}%{January 10, 2019}
\accepted{}%{\today}
\submitjournal{ApJ}

\shorttitle{ALMA observations of the GMC in M33}
\shortauthors{Kondo et al.}

\begin{document}

\title{ALMA Observations of Giant Molecular Clouds in M33. III.: \\Spatially Resolved Features of the Star-Formation Inactive Million-solar-mass Cloud}

\correspondingauthor{Hiroshi Kondo, Kazuki Tokuda}
\email{konkon.5164@gmail.com; tokuda@p.s.osakafu-u.ac.jp}

\author{Hiroshi Kondo}
\affiliation{Department of Physical Science, Graduate School of Science, Osaka Prefecture University, 1-1 Gakuen-cho, Naka-ku, Sakai, Osaka 599-8531, Japan}

\author[0000-0002-2062-1600]{Kazuki Tokuda}
\affiliation{Department of Physical Science, Graduate School of Science, Osaka Prefecture University, 1-1 Gakuen-cho, Naka-ku, Sakai, Osaka 599-8531, Japan}
\affiliation{National Astronomical Observatory of Japan, National Institutes of Natural Science, 2-21-1 Osawa, Mitaka, Tokyo 181-8588, Japan}

\author{Kazuyuki Muraoka}
\affiliation{Department of Physical Science, Graduate School of Science, Osaka Prefecture University, 1-1 Gakuen-cho, Naka-ku, Sakai, Osaka 599-8531, Japan}

\author{Atsushi Nishimura}
\affiliation{Department of Physical Science, Graduate School of Science, Osaka Prefecture University, 1-1 Gakuen-cho, Naka-ku, Sakai, Osaka 599-8531, Japan}

\author{Shinji Fujita}
\affiliation{Department of Physical Science, Graduate School of Science, Osaka Prefecture University, 1-1 Gakuen-cho, Naka-ku, Sakai, Osaka 599-8531, Japan}

\author{Tomoka Tosaki}
\affiliation{Joetsu University of Education, Yamayashiki-machi, Joetsu, Niigata 943-8512, Japan}

\author{Sarolta Zahorecz}
\affiliation{Department of Physical Science, Graduate School of Science, Osaka Prefecture University, 1-1 Gakuen-cho, Naka-ku, Sakai, Osaka 599-8531, Japan}
\affiliation{National Astronomical Observatory of Japan, National Institutes of Natural Science, 2-21-1 Osawa, Mitaka, Tokyo 181-8588, Japan}

\author{Rie E. Miura}
\affiliation{National Astronomical Observatory of Japan, National Institutes of Natural Science, 2-21-1 Osawa, Mitaka, Tokyo 181-8588, Japan}

\author{Masato I.N. Kobayashi}
\affiliation{Astronomical Institute, Tohoku University, 6-3, Aramaki, Aoba-ku, Sendai, Miyagi 980-8578, Japan}

\author{Sachiko Onodera}
\affiliation{Meisei University, 2-1-1 Hodokubo, Hino, Tokyo 191-0042, Japan}

\author{Kazufumi Torii}
\affiliation{Nobeyama Radio Observatory, National Astronomical Observatory of Japan (NAOJ), National Institutes of Natural Sciences (NINS), 462-2 Nobeyama, Minamimaki, Minamisaku-gun, Nagano 384-1305, Japan}

\author{Nario Kuno}
\affiliation{Department of Physics, Graduate School of Pure and Applied Sciences, University of Tsukuba, 1-1-1 Tennodai, Tsukuba, Ibaraki 305-8577, Japan}
\affiliation{Tomonaga Center for the History of the Universe, University of Tsukuba, Tsukuba, Ibaraki 305-8571, Japan}

\author{Hidetoshi Sano}
\affiliation{National Astronomical Observatory of Japan, National Institutes of Natural Science, 2-21-1 Osawa, Mitaka, Tokyo 181-8588, Japan}

\author[0000-0001-7826-3837]{Toshikazu Onishi}
\affiliation{Department of Physical Science, Graduate School of Science, Osaka Prefecture University, 1-1 Gakuen-cho, Naka-ku, Sakai, Osaka 599-8531, Japan}

\author{Kazuya Saigo}
\affiliation{National Astronomical Observatory of Japan, National Institutes of Natural Science, 2-21-1 Osawa, Mitaka, Tokyo 181-8588, Japan}

\author{Yasuo Fukui}
\affiliation{Department of Physics, Nagoya University, Chikusa-ku, Nagoya 464-8602, Japan}
\affiliation{Institute for Advanced Research, Nagoya University, Furo-cho, Chikusa-ku, Nagoya 464-8601, Japan}

\author{Akiko Kawamura}
\affiliation{National Astronomical Observatory of Japan, National Institutes of Natural Science, 2-21-1 Osawa, Mitaka, Tokyo 181-8588, Japan}

\author{Kisetsu Tsuge}
\affiliation{Department of Physics, Nagoya University, Chikusa-ku, Nagoya 464-8602, Japan}

\author{Kengo Tachihara}
\affiliation{Department of Physics, Nagoya University, Chikusa-ku, Nagoya 464-8602, Japan}

\begin{abstract}

We present $^{12}$CO ($J$\,=\,2--1), $^{13}$CO ($J$\,=\,2--1), and C$^{18}$O ($J$\,=\,2--1) observations toward GMC-8, one of the most massive giant molecular clouds (GMCs) in M33 using ALMA with an angular resolution of 0\farcs44\,$\times$\,0\farcs27 ($\sim$2\,pc\,$\times$\,1\,pc). 
The earlier studies revealed that its high-mass star formation is inactive in spite of a sufficient molecular reservoir with a total mass of $\sim$10$^{6}$\,$M_{\odot}$.
The high-angular resolution data enable us to resolve this peculiar source down to a molecular clump scale. One of the GMC's remarkable features is that a round-shaped gas structure (the $``$Main cloud$"$) extends over the $\sim$50\,pc scale, which is quite different from the other two active star-forming GMCs dominated by remarkable filaments/shells obtained by our series of studies in M33. The fraction of the relatively dense gas traced by the $^{13}$CO data with respect to the total molecular mass is only $\sim$2\%, suggesting that their spatial structure and the density are not well developed to reach an active star formation. The CO velocity analysis shows that the GMC is composed of a single component as a whole, but we found some local velocity fluctuations in the Main cloud and extra blueshifted components at the outer regions. 
Comparing the CO with previously published large-scale H$\;${\sc i} data, we suggest that an external atomic gas flow supplied a sufficient amount of material to grow the GMC up to $\sim$10$^6$\,$M_{\odot}$.
\end{abstract}


\section{Introduction} \label{sec:intro}
Giant molecular clouds (GMCs) are the fundamental cradles of star formation (see the review by e.g., \citealt{Fukui10}, and \citealt{Heyer15}). The formed stars, especially high-mass stars, eventually destroy and deform the parental molecular clouds, and provide metal-rich ingredients; thus the lifecycle likely controls the galaxy evolution itself. 
A wide-field CO imaging along the Milky Way (MW, e.g., \citealt{Dame01,Jackson06,Nishimura15,Umemoto17,Su19}) is a powerful tool to fully resolve GMCs spatially. However, investigating individual targets alone provides us only a snapshot of the evolutionary sequence, and thus comprehensive studies throughout a galaxy outside the Milky Way, which does not suffer from the contamination problem at the line of sight and the distance ambiguity, have great advantages to compile a large number of GMC sample uniformly. 

Some of the comprehensive CO surveys toward the Local group of galaxies, especially the Large Magellanic Cloud (LMC), revealed a GMC scale resolution ($\sim$40\,pc) view and derived an evolutionary track of the GMC \citep{Fukui99,Fukui08,Kawamura09}. One striking result in the LMC is that a large fraction ($\sim$1/4) of the GMCs does not harbor any bright H$\;${\sc ii} regions, indicating that the evolutionary stage is likely in an early phase prior to high-mass star formation (hereafter, starless GMC). These starless GMCs are supposed not to be exposed to any stellar feedback and thus they are suitable targets to investigate the initial condition of not only star formation but also molecular cloud formation. 

Observational studies in the ALMA era enabled us to reveal the substructure of molecular clouds in the Magellanic Clouds \citep[e.g.,][]{Indebetouw13,Fukui15,Fukui19,Muraoka17,Saigo17,Naslim18,Sawada18,Tokuda19,Wong19} with a size scale of $\sim$0.1--100\,pc. The pilot CO studies toward individual targets found that the active star-forming regions show well-developed structures, such as cores/filaments, whereas some of the inactive star-forming regions show a diffuse/extended feature \citep{Sawada18}. 

More distant galaxies are also vital targets to comprehensively understand the picture of star/molecular cloud formation. 
Although the achievable spatial resolution is more than 10 times coarser than that in the Magellanic clouds, the Triangulum Galaxy (M33) provides us with a unique laboratory located nearest to us, $\sim$840\,kpc \citep{Freedman01}, among similar targets accessible from the ALMA site. 
The flocculent spiral galaxy, M33, has the potential to investigate the effects of the complex gas dynamics on star formation and galaxy evolution \citep{Wada11}, which may not be achieved by studies of $``$grand design$"$ spiral galaxies, such as M51. Our M33 observations with the ALMA found giant molecular filaments whose length is more than 50\,pc along one of the stellar spiral arm (\citealt{Tokuda20}, hereafter Paper~I).
In addition to this, there is the highly active star-forming GMC at a vicinity of NGC~604, which contains the largest H$\;${\sc ii} region in the Local Group of galaxies, enabling us to study such an extreme star formation (\citealt{Muraoka20}, hereafter, Paper~II). 

\begin{figure}[htbp]
\begin{center}
\includegraphics[width=150mm]{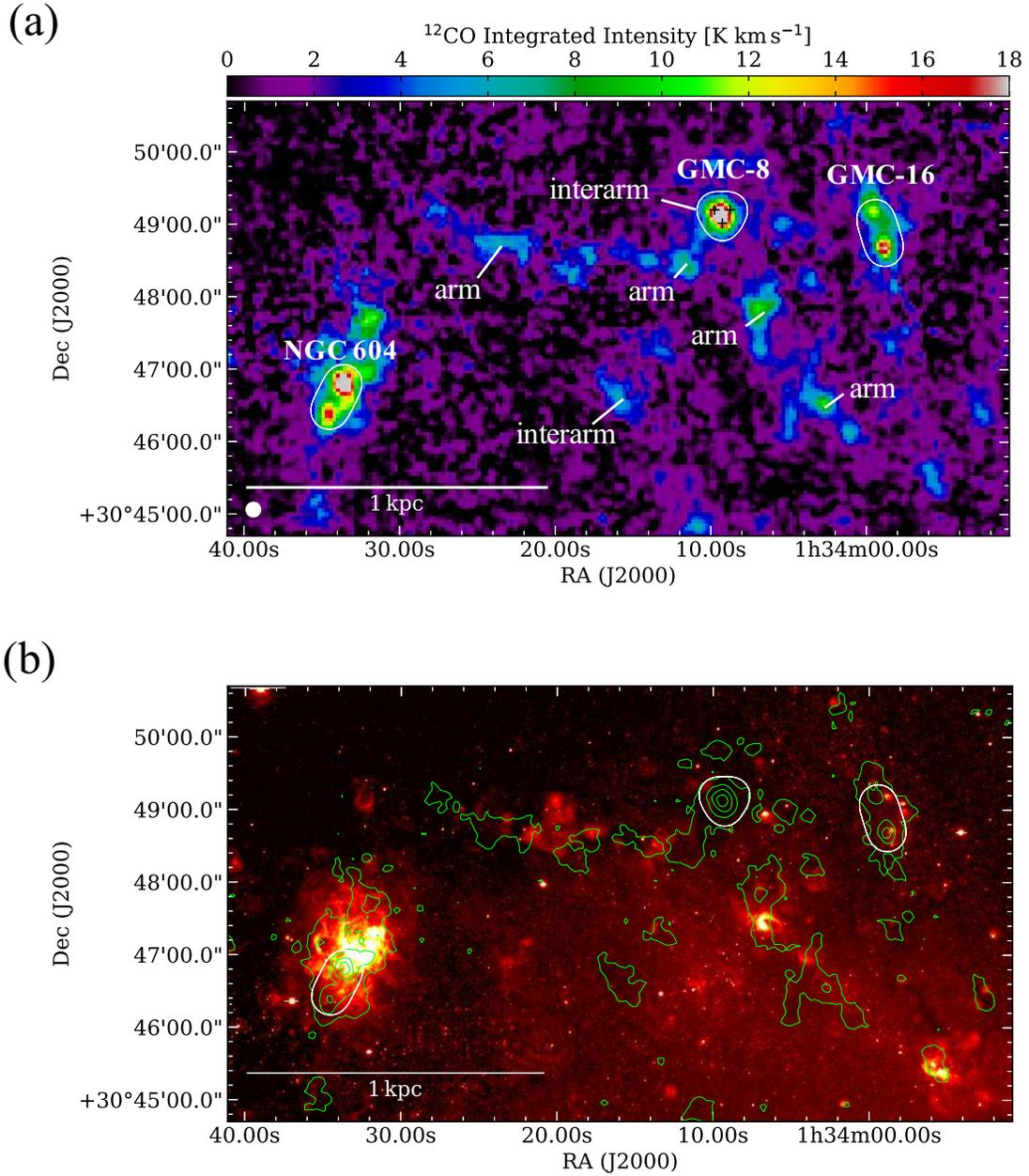}
\end{center}
\caption{Distributions of molecular and ionized gas toward the northern part of M33. (a) The color-scale image shows the integrated intensity image of $^{12}$CO\,($J$\,=\,2--1) with the IRAM 30\,m telescope \citep{Druard14}. The labels of $``$arm$"$ and $``$interarm$"$ represent GMC locations with respect to the spiral arm categorized by \cite{Roso07}. The white lines show the field coverage of our ALMA studies (see \citealt{Tokuda20} for GMC-16, and \citealt{Muraoka20} for NGC~604). The black crosses in the GMC-8 field denote the central coordinates in each pointing, ($\alpha_{\rm J2000.0}$, $\delta_{\rm J2000.0}$) = 
(1$^{\rm h}$34$^{\rm m}$09\fs27, +30\arcdeg49\arcsec01\farcs2), 
(1$^{\rm h}$34$^{\rm m}$08\fs77, +30\arcdeg49\arcsec12\farcs4), and 
(1$^{\rm h}$34$^{\rm m}$09\fs77, +30\arcdeg49\arcsec12\farcs4).
The white circle at the lower left corner represents the angular resolution of the CO image, $\sim$11\arcsec. (b) The heat-map shows the H$\alpha$ emission \citep{Hoopes00}. The green contours show the CO image, which is the same as panel (a). The lowest contour and subsequent steps are 3\,K\,km\,s$^{-1}$ and 6\,K\,km\,s$^{-1}$, respectively.
\label{fig:IRAM}}
\end{figure}

Our target object in this study is GMC-8 (Figure\,\ref{fig:IRAM}) cataloged by the ASTE $^{12}$CO\,(3--2) survey of \cite{Miura12}. The GMC corresponds to the cloud number of 245 from the IRAM $^{12}$CO\,(2--1) survey by \cite{Gratier12} (see also the $^{12}$CO\,(1--0) observations, \citealt{Roso07,Onodera10,Tosaki11}). This cloud has the strongest emission in $^{12}$CO\,(1--0), inferring that GMC-8 is one of the most massive among more than 250 GMCs in M33. The molecular mass of GMC-8 is $\sim$10$^{6}$\,$M_{\odot}$ based on the $^{12}$CO luminosity. \cite{Miura12} reported that the star-formation activity of this cloud is inactive, indicating that the GMC is presumably an exception in the galactic-scale perspective, i.e., the Kennicutt–Schmidt law \citep{Schmidt59,Kennicutt98,Kennicutt12}. 
The previous infrared/optical observations found 24\,$\mu$m \citep{Verley07} and H$\alpha$ \citep{Hodge99} sources toward this GMC (see also Section zr\ref{result:space}), but \cite{Gratier12} did not categorize GMC-8 as an exposed star formation phase based on their extreme faintness feature compared to the other active star-forming GMCs. \cite{Roso07} labeled its location as the interarm region of the northern M33 (see Figure\,\ref{fig:IRAM}), and thus the environment may be relevant to the inactive nature of star formation. Such a starless GMC is highly rare in the Milky Way. A few examples in the solar neighborhood are the Maddalena's cloud \citep{Maddalena85} and the Cygnus~OB7 GMC \citep{Dobashi94,Dobashi96}, which are less massive than GMC-8 (see also Sect.~\ref{dis:LoGMC}). Numerical simulations of GMC formation suggest that the converging flow of H$\;${\sc i} gas \citep{Hennebelle99,Koyama04,Inoue09} makes a large amount of molecular materials whose mass is $\sim$10$^{4}$\,$M_{\odot}$ \citep{InoueInu12}; meanwhile, the typical cross section in these simulations is 10\,pc by 10\,pc, which suggests that drastic flows with 100\,pc scales are required to be 10$^6$\,$M_{\odot}$. Semianalytic studies of the GMC mass function \citep[e.g.,][]{Kobayashi18} also suggest that it is not easy to produce more than $\sim$10$^{5}$\,$M_{\odot}$ clouds within a GMC lifetime via converging flows with a typical condition. Therefore GMC-8 in M33 is a vital target to explore the formation mechanism of massive GMCs.

This paper presents our new ALMA CO data of GMC-8 (Sect.\,\ref{sec:obs}) with a spatial resolution of $\sim$1\,pc and investigates its resolved physical properties (Sect.\,\ref{sec:results}). 
In this study, we discuss the formation of GMC-8 as a very massive GMC by combining the previously obtained large-scale CO and H$\;${\sc i} maps (Sect.\,\ref{sec:dis}).

%\if0
%\fi

\section{The Data} \label{sec:obs}

We conducted ALMA Band 6 (211--275\,GHz) observations toward three GMCs in M33 using the ALMA 12 and the 7\,m array of the Atacama Compact Array (ACA) in Cycle\,5 (P.I.: K.Muraoka \#2017.1.00461.S). The observation details and data reduction process were described in Paper~I. Here we briefly summarize the observation settings and the data qualities for GMC-8. Three pointing observations at the central coordinates of ($\alpha_{\rm J2000.0}$, $\delta_{\rm J2000.0}$) = 
(1$^{\rm h}$34$^{\rm m}$09\fs27, +30\arcdeg49\arcsec01\farcs2), 
(1$^{\rm h}$34$^{\rm m}$08\fs77, +30\arcdeg49\arcsec12\farcs4), and 
(1$^{\rm h}$34$^{\rm m}$09\fs77, +30\arcdeg49\arcsec12\farcs4) made a triangular-shaped coverage, as shown in Figure \ref{fig:IRAM}. The target molecular lines were $^{12}$CO\,($J$\,=\,2--1), $^{13}$CO\,($J$\,=\,2--1), and C$^{18}$O\,($J$\,=\,2--1). The wavelength of the continuum band is 1.3~mm with a sensitivity of $\sim$0.02\,mJy\,beam$^{-1}$. 

We combined the 12\,m and 7\,m array data with the \texttt{feathering} task. The beam size of the $^{12}$CO combined data is 0\farcs42 $\times$ 0\farcs26  (1.7\,pc $\times$ 1.0\,pc), and the rms noise level is $\sim$0.9\,K at a velocity resolution of $\sim$0.2\,km s$^{-1}$. We estimated the missing flux of the $^{12}$CO ALMA data in GMC-8 by comparing with the single-dish IRAM 30\,m data \citep{Druard14}. The total missing flux across the observed field is $\sim$30\%, and thus we additionally combined the ALMA and the IRAM data using the feathering technique. We note that because the velocity resolution of the IRAM data, 2.6\,km\,s$^{-1}$, is much coarser than that of the ALMA data, we use the ALMA data alone for analyses that require a high-velocity resolution (see the captions in each figure). 

We made moment-masked cube data \citep[e.g.,][]{Dame11,Nishimura15} to suppress the noise effect. We set the emission-free pixels, which are determined by significant emission from the smoothed data in the velocity/spatial axes, as zero values. We used the processed data to make moment maps of $^{12}$CO and $^{13}$CO.

\section{Results} \label{sec:results}

\subsection{Spatial Distributions and Physical Properties of GMC-8 in $^{12}$CO and $^{13}$CO}
\label{result:space}
\begin{figure}[htbp]
\includegraphics[width=180mm]{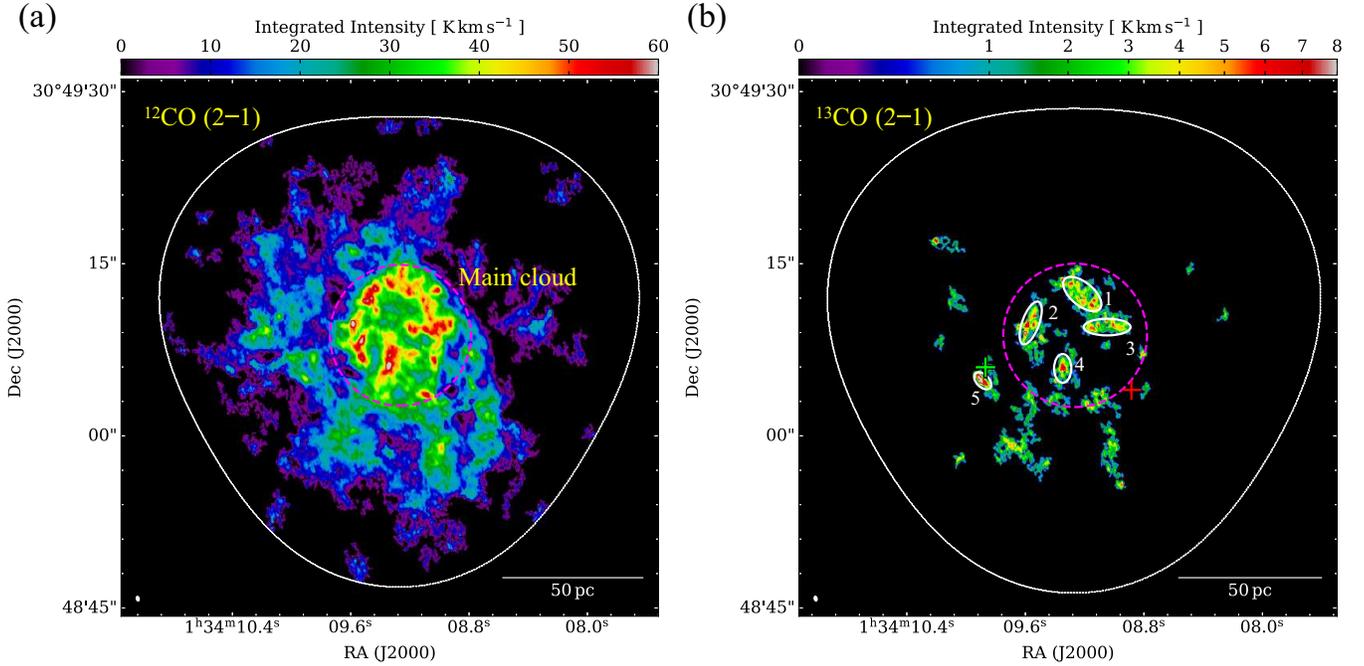}
\caption{Integrated intensity map of $^{12}$CO\,($J$\,=\,2--1) and $^{13}$CO\,($J$\,=\,2--1) in M33-GMC-8. (a) Color-scale image shows the velocity-integrated intensity map of $^{12}$CO\,($J$\,=\,2--1) combining the ALMA data (the 12\,m + 7\,m array) with IRAM-30\,m data. The angular resolution is given by the white ellipse in the lower left corner.
The magenta dashed circle shows the arbitrarily defined region of the Main cloud (see the text). The white line shows the field coverage of the ALMA observations. (b) Same as (a) but for the $^{13}$CO\,($J$\,=\,2--1) image obtained by the ALMA 12\,m + 7\,m array. The green and red crosses represent the position of the 24\,$\mu$m \citep{Verley07} and H$\alpha$ source \citep{Hodge99}, respectively. White ellipses show the sizes of the selected $^{13}$CO clumps (see Table~\ref{tab:13COclump}) determined by 2D Gaussian fitting.
\label{fig:12CO13CO}}
\end{figure}

Figure\,\ref{fig:12CO13CO} shows the velocity-integrated intensity map (moment\,0) of $^{12}$CO($J$\,=\,2--1) and $^{13}$CO($J$\,=\,2--1). One of the remarkable features in the $^{12}$CO map (panel(a)) is a widely extended and round-shaped structure with a diameter of $\sim$50\,pc as indicated by the magenta dashed circle in the figure. We call this structure as the $``$Main cloud$"$ hereafter. It has a relatively sharp boundary at the western side, although it extends outward with several arm-like features in the other directions. The peak integrated intensity is $\sim$60\,K\,km\,s$^{-1}$, corresponding to an H$_2$ column density of $\sim$2\,$\times$\,10$^{22}$\,cm$^{-2}$ if we assume a constant $^{12}$CO($J$\,=\,2--1)/$^{12}$CO($J$\,=\,1--0) ratio of $\sim$0.5, obtained from the single-dish data \citep{Tosaki11,Druard14}, and the $X_{\rm CO}$ factor of 4\,$\times$\,10$^{20}$\,cm$^{-2}$\,(K\,km\,s$^{-1}$)$^{-1}$ \citep{Druard14}. The total mass integrated within the observed field (hereafter $M_{\rm ^{12}CO}$) is $\sim$4$\times$10$^{6}$\,$M_{\odot}$, which is consistent with the previous studies \citep{Druard14}.

In contrast to $^{12}$CO, the $^{13}$CO moment\,0 show a clumpy distribution throughout the observed field. As one can see, there are $\sim$20 spatially isolated components, and we call these objects as $``^{13}$CO clumps.$"$ The intensities of the $^{13}$CO clumps are $\sim$2--10\,K\,km\,s$^{-1}$, which are at most one-tenth lower than that of the two active star-forming GMCs (Paper~I for GMC-16, and Paper~II for NGC~604 GMC). We selected five bright clumps whose peak $^{13}$CO intensity exceeds 5\,K\,km\,s$^{-1}$, corresponding to the column density of $\sim$1.5$\times$10$^{21}$\,cm$^{-2}$, as as indicated in Figure\,\ref{fig:12CO13CO}(b).
We derived the size of the $^{13}$CO clumps by 2D Gaussian fittings at each clump in the image plane within the regions at a more than 3$\sigma$ contour. Table~\ref{tab:13COclump} gives the physical properties of the $^{13}$CO clumps.
The intensity and typical sizes are similar to Galactic low-mass star-forming regions, such as Taurus, Chamaeleon, and Aquila \citep{Mizuno95,Mizuno99,Kawamura99}, and a Planck Cold Cloud in the LMC \citep{Wong17}, traced by their $^{13}$CO observations with a similar angular resolution. To characterize these clumps, we estimated the H$_2$ column density assuming the local thermodynamic equilibrium with a uniform excitation temperature of $\sim$10\,K and a relative [H$_2$]/[$^{13}$CO] abundance of 1.4\,$\times$\,10$^{6}$ (see also Paper~I, II). The mean column densities and masses ($M_{\rm LTE}$) are $\sim$(3--5)\,$\times$\,10$^{21}$\,cm$^{-2}$ and $\sim$(2--10)\,$\times$\,10$^{3}$\,$M_{\odot}$. 

%\if0
\begin{deluxetable*}{cccccccccc}
\tabletypesize{\scriptsize}
\tablewidth{0pt} 
%\tablenum{1}
\tablecaption{Properties in Each $^{13}$CO Clump
\label{tab:13COclump}}
\tablehead{
\colhead{Source ID}      & \colhead{R.A.}               & \colhead{Decl.}              & \colhead{Size}         &   \colhead{$v_{\rm center}$}${(1)}$  &
\colhead{$v_{\rm FWHM}$}${(2)}$ & \colhead{$N_{\rm H_2}^{\rm mean}$}${(3)}$ & \colhead{$N_{\rm H_2}^{\rm peak}$}${(4)}$ &\colhead{Virial Mass}${(5)}$   & \colhead{$M_{\rm LTE}$}${(6)}$
\\[-1.0mm]
      & (J2000.0)  & (J2000.0) & (pc)   & (km\,s$^{-1}$) & (km\,s$^{-1}$)  &(cm$^{-2}$) &(cm$^{-2}$) &($M_{\odot}$)& ($M_{\odot}$) %\\[-1.0mm]
}
%\colnumbers
\startdata 
clump-1 & 01 34 09.22 & 30 49 12.19 &  20$\times$5.8 & $-$247.7 & 5.7 & 3$\times$10$^{21}$ & 1$\times$10$^{22}$ & 4$\times$10$^{4}$ & 1$\times$10$^{4}$ \\
clump-2 & 01 34 09.57 & 30 49 09.72 & 9.1$\times$3.3 & $-$247.8 & 5.1 & 4$\times$10$^{21}$ & 1$\times$10$^{22}$ & 2$\times$10$^{4}$ & 4$\times$10$^{3}$ \\
clump-3 & 01 34 09.06 & 30 49 09.76 &  18$\times$8.5 & $-$242.4 & 4.7 & 3$\times$10$^{21}$ & 8$\times$10$^{21}$ & 3$\times$10$^{4}$ & 5$\times$10$^{3}$ \\
clump-4 & 01 34 09.34 & 30 49 05.86 & 7.3$\times$3.3 & $-$257.3 & 4.4 & 4$\times$10$^{21}$ & 9$\times$10$^{21}$ & 1$\times$10$^{4}$ & 2$\times$10$^{3}$ \\
clump-5 & 01 34 09.89 & 30 49 04.82 & 5.4$\times$1.8 & $-$251.7 & 2.3 & 5$\times$10$^{21}$ & 2$\times$10$^{22}$ & 2$\times$10$^{3}$ & 3$\times$10$^{3}$ \\
\enddata
\tablecomments{(1) Central velocity. (2) Velocity width of $^{13}$CO($J$ = 2--1) emission. (3) Mean of H$_2$ column density. (4) Peak of H$_2$ column density. (5) Virial mass calculated from the following formula, $M_{\rm vir}$ = $210 \Delta V^2 R$, where $R$ is the geometric mean of the source size. (6) Total LTE mass within each clump.}
\end{deluxetable*}
%\fi

\begin{figure}[htbp]
\begin{center}
\includegraphics[width=90mm]{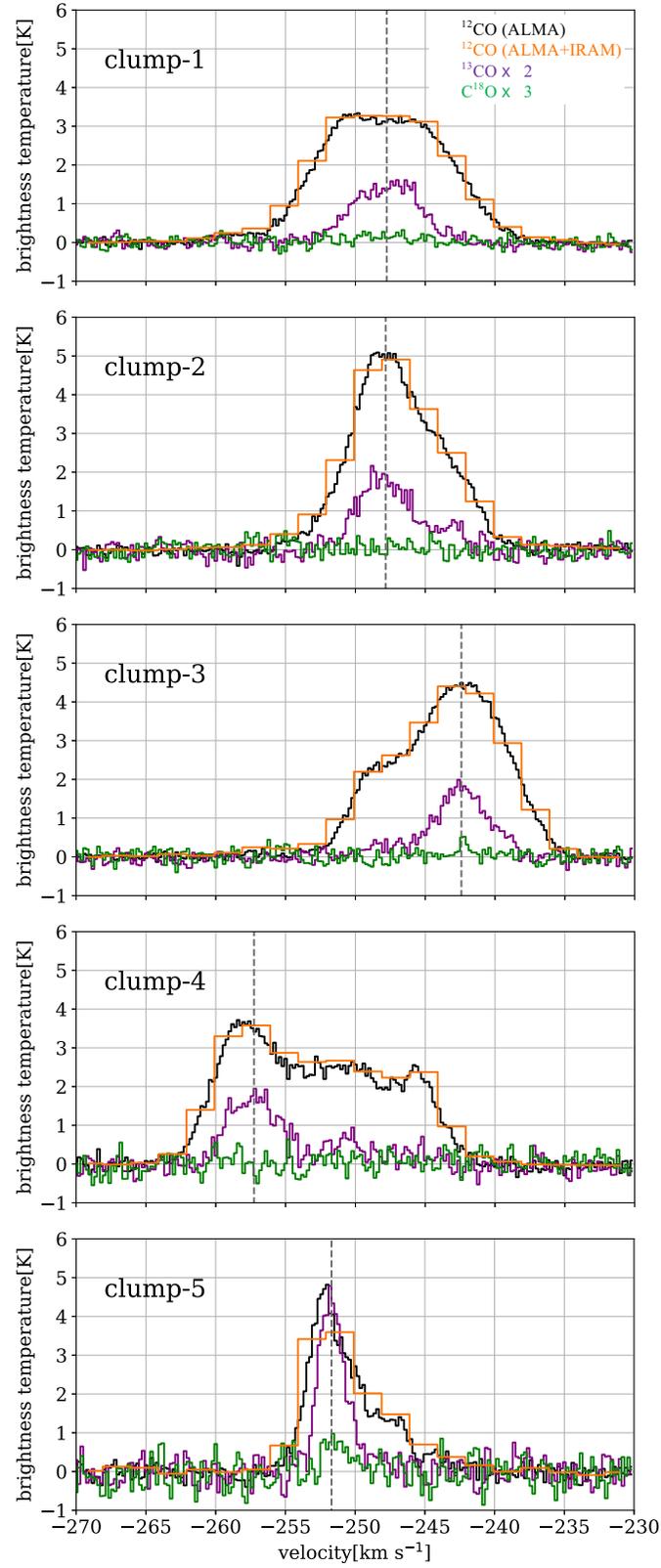}
\end{center}
\caption{Average spectra of $^{12}$CO, $^{13}$CO, and C$^{18}$O($J$\,=\,2--1) in the $^{13}$CO clumps. The gray dashed line indicates the central velocities determined by fitting the $^{13}$CO spectra with a single Gaussian profile.
\label{fig:13CO_clump}}
\end{figure}

For each selected $^{13}$CO clump, we performed the average scheme within the regions showing significant $^{13}$CO emission above 3$\sigma$ to obtain the representative spectra. Figure~\ref{fig:13CO_clump} show the $^{12}$CO, $^{13}$CO and C$^{18}$O spectra. The subsequent Gaussian fitting to the $^{13}$CO spectra determined their central velocity ($v_{\rm center}$) and velocity width in FWHM ($v_{\rm FWHM}$).
The $v_{\rm center}$ has a large variety from $-$257\,km\,s$^{-1}$ to $-$242\,km\,s$^{-1}$. Although the $^{13}$CO profiles have a Gaussian shape, some $^{12}$CO spectra show complex profiles, which likely have multiple velocity components. The averaged analysis allows us to search weak C$^{18}$O emission at the clumps. At clump-3, we found significant C$^{18}$O emission, whose peak velocity is similar to that of $^{13}$CO. The detection of C$^{18}$O with more than 3$\sigma$ noise level of the average spectra implies the presence of dense molecular materials with the density of $\sim$10$^{4}$\,cm$^{-3}$. At clump-5, we found marginal C$^{18}$O emission and the highest $^{13}$CO integrated intensity in the observed field (see Figure~\ref{fig:12CO13CO}). The (column) density peak presents at the region close to the GMC edge instead of an inner side of the Main cloud. 

The integrated total $M_{\rm LTE}$ in the observed field (hereafter $M_{\rm ^{13}CO}$) is $\sim$6\,$\times$\,10$^4$\,$M_{\odot}$. The mass ratio ($M_{\rm ^{13}CO}$/$M_{\rm ^{12}CO}$) is $\sim$2\%, which is remarkably lower than that of the other star-forming GMCs in M33 (Table\,\ref{tab:m33gmc}, see also Paper~I, II) and the Galactic plane average of $\sim$24\% \citep{Torii19}. This result shows that the relatively dense gas fraction with respect to the total molecular gas is not high compared to the other targets, and consistent with nondetection of 1.3~mm and C$^{18}$O emission tracing $\gtrsim$10$^4$\,cm$^{-3}$ dense gas at this resolution/sensitivity. 
%\if0
\begin{deluxetable*}{cccccc}
\tabletypesize{\scriptsize}
\tablewidth{0pt} 
%\tablenum{2}
\tablecaption{The Comparison of Dense Gas Properties Among the Three GMC in M33
\label{tab:m33gmc}}
\tablehead{
\colhead{GMC Name} & \colhead{C$^{18}$O Detection}  & \colhead{1.3\,mm Detection}　& \colhead{$M_{\rm ^{12}CO}$}${(1)}$ & \colhead{$M_{\rm ^{13}CO}$}${(2)}$  & \colhead{$M_{\rm ^{13}CO}$/$M_{\rm ^{12}CO}$}   
\\[-1.0mm]
      &    &                  & ($M_{\odot}$)           & ($M_{\odot}$)            &  
}
%\colnumbers
\startdata 
GMC-16      & Y         & Y  & $\sim$2\,$\times$10$^6$ & $\sim$2\,$\times$10$^5$ & $\sim$0.1 \\
NGC~604 GMC & Y         & Y  & $\sim$3\,$\times$10$^6$ & $\sim$3\,$\times$10$^5$ & $\sim$0.1 \\
GMC-8       & N${(3)}$ & N  & $\sim$4\,$\times$10$^6$ & $\sim$6\,$\times$10$^4$ & $\sim$0.02 \\
\enddata
\tablecomments{(1) Total molecular gas mass derived from the $^{12}$CO ($J$\,=\,2--1) luminosity assuming the $X_{\rm CO}$ factor of 4\,$\times$10$^{20}$\,cm$^{-2}$\,(K\,km\,s$^{-1}$)$^{-1}$. Based on the single-dish data, we applied $^{12}$CO ($J$\,=\,2--1)/$^{12}$CO ($J$\,=\,1--0) ratios of 0.7 for GMC-16, 0.85 for NGC~604 and 0.5 for GMC-8. (2) LTE mass derived from the $^{13}$CO ($J$\,=\,2--1) data. (3) Smoothing analysis toward some of the selected $^{13}$CO clumps shows significant C$^{18}$O emission (see the text).}
\end{deluxetable*}
%\fi

\begin{figure}[htbp]
\includegraphics[width=180mm]{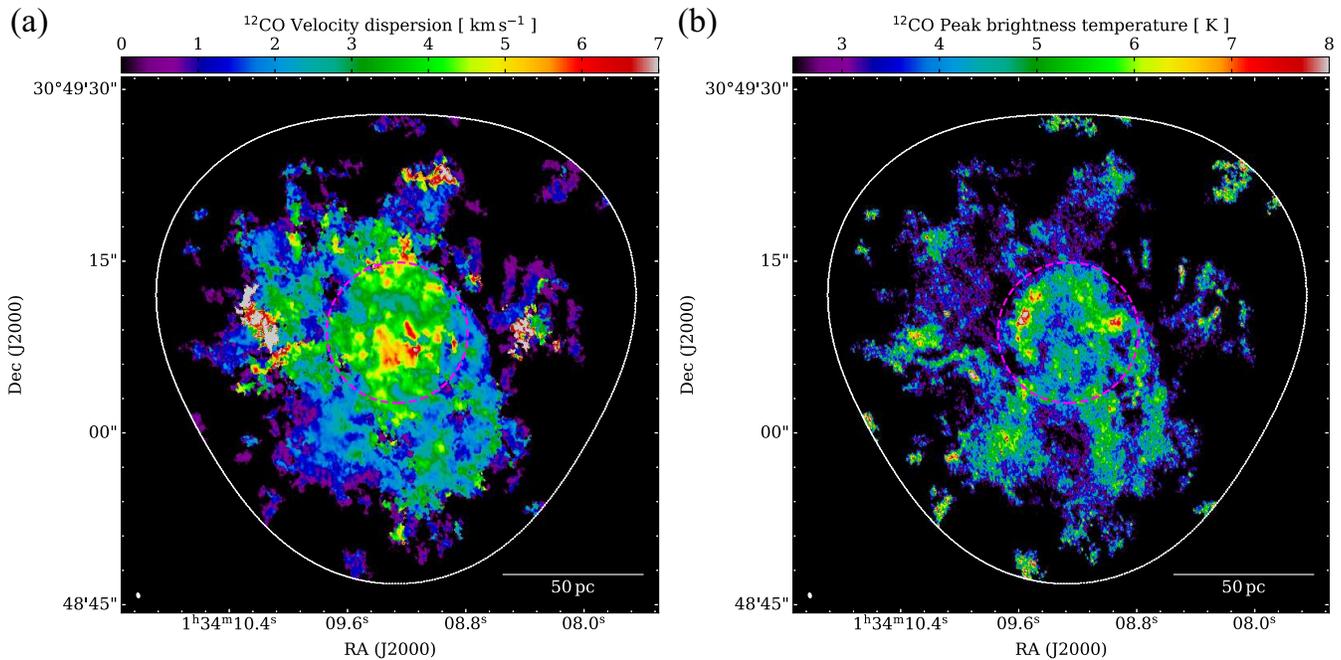}
\caption{(a) Color-scale image shows the second-moment map of the $^{12}$CO ($J$\,=\,2--1) data with the 12\,m + 7\,m array. Note that the color scale at the pixels above 7\,km\,s$^{-1}$ is saturated.} The magenta dashed circle is the same as that in Figure\,\ref{fig:12CO13CO}. The white ellipse in the lower left corner shows the angular resolution. The white line shows the field coverage of the ALMA observations.
(b) Same as (a) but for the peak brightness temperature map of the $^{12}$CO ($J$\,=\,2--1) data. 
\label{fig:mom12}
\end{figure}

Figure\,\ref{fig:mom12} shows the velocity dispersion, $\sigma_v$ (moment\,2), map and the peak brightness temperature map in $^{12}$CO. 
The GMC center shows a higher velocity dispersion than the outer parts. In the Main cloud, the FWHM line width of the average spectra is $\sim$11\,km\,s$^{-1}$ derived by the Gaussian fitting. We will describe the velocity structure further in the next subsection (Sect.~\ref{result:veloicty}). The peak brightness temperature (panel (b)) ranges from a few to 8\,K, suggesting that the excitation temperature is low ($\lesssim$10\,K). In contrast to this, the most active star-forming complex, NGC~604, shows much higher temperature of $\sim$40\,K (Paper~II), indicating that GMC-8 currently does not harbor any remarkable heating sources. Although the edge of the Main cloud has a higher brightness temperature in the field, these regions are apart from the faint 24\,$\mu$m and H$\alpha$ sources (Figure\,\ref{fig:12CO13CO}(b)). Note that the position accuracy of the two data sets, 3\arcsec for 24\,$\mu$m \citep{Rieke04} and $\sim$0\farcs5 for H$\alpha$ (see \citealt{Hoopes00,Bailer2018}), does indicate that they are outside of the Main cloud.

\subsection{Velocity Structures of GMC-8} \label{result:veloicty}

\begin{figure}[htbp]
\includegraphics[width=180mm]{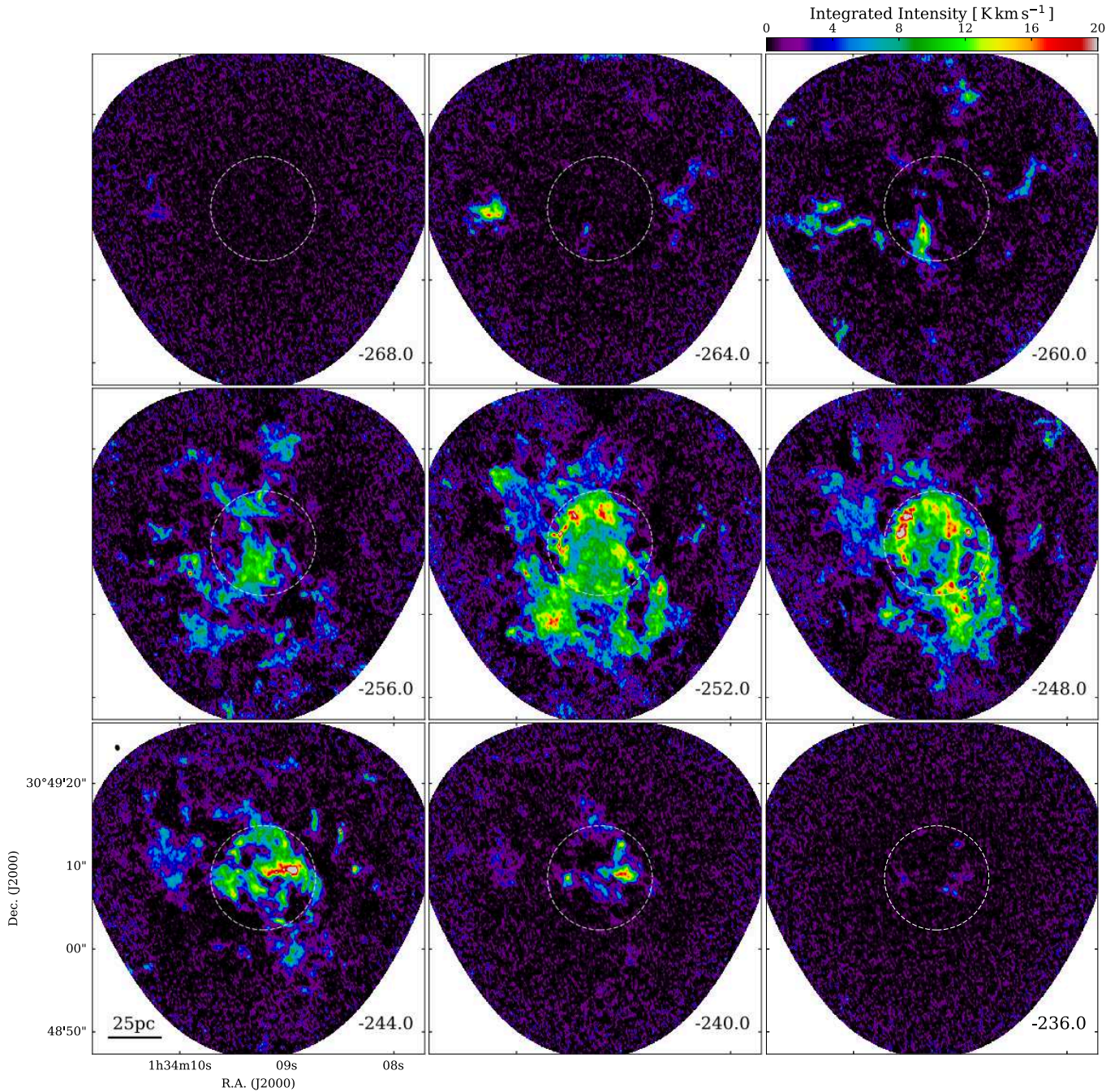}
\caption{Velocity channel maps toward GMC-8 in $^{12}$CO (the 12\,m + 7\,m array data). The black ellipse in the upper left corner of the lower left panel shows the angular resolution. The lowest velocities in the units of km\,s$^{-1}$ are shown in the lower right corners of each panel. The white dashed circle shows the arbitrarily defined region of the Main cloud (see the text).
\label{fig:chanmap}}
\end{figure}

\begin{figure}[htbp]
\begin{center}
\includegraphics[width=120mm]{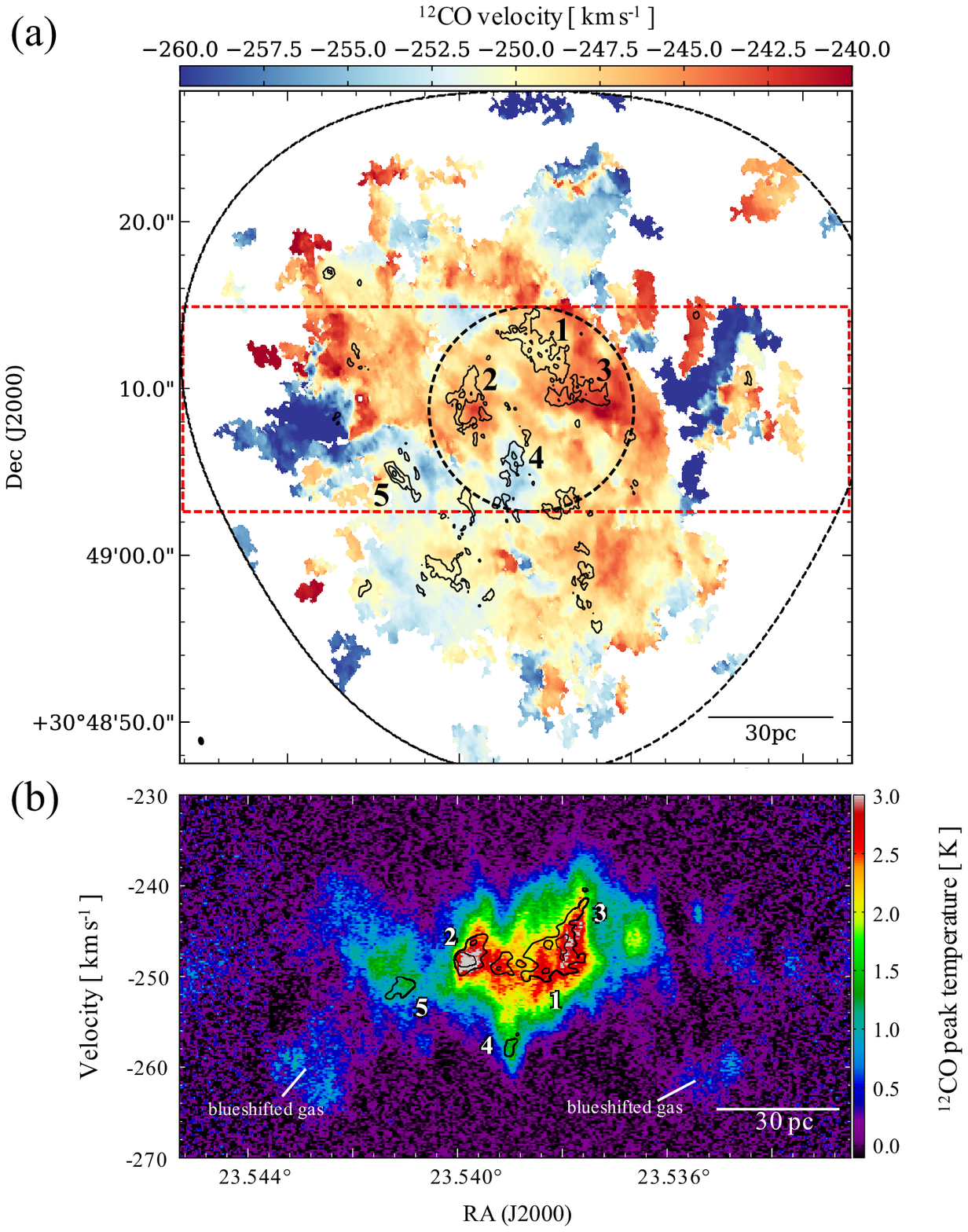}
\end{center}
\caption{(a) Color-scale image shows the first-moment map of the $^{12}$CO ($J$\,=\,2--1) data with the 12\,m + 7\,m array. The contour shows the $^{13}$CO integrated intensity map. 
The lowest contour and subsequent steps are 2\,K\,km\,s$^{-1}$ and 4\,K\,km\,s$^{-1}$, respectively. The numbers show IDs of $^{13}$CO clumps. The black ellipse in the lower left corner shows the angular resolution. 
(b) A $^{12}$CO ($J$\,=\,2--1) position-velocity diagram along the R.A. axis, shown by the red dashed rectangle in panel (a). The contour shows a smoothed $^{13}$CO image with an angular resolution of 0\farcs88\,$\times$\,0\farcs53 and a velocity resolution of 0.2\,km\,s$^{-1}$. The contour levels are 0.22 and 0.32\,K. Each number shows the IDs of the $^{13}$CO clumps. 
\label{fig:mom1_pv}}
\end{figure}

The velocity channel maps of $^{12}$CO in Figure\,\ref{fig:chanmap} show that the emission are continuously distributed over the velocity range of $-$268\,--\,$-$236\,km\,s$^{-1}$. Figure\,\ref{fig:mom1_pv} shows the intensity-weighted velocity (moment1) map. Toward the Main cloud, panel (a) shows that there is no velocity jump larger than $\sim$5\,km\,s$^{-1}$ (see also panel (b)).
If there are more than two velocity components, the velocity maps show opposite results, i.e., a drastic velocity gap in moment\,1 and a large dispersion in moment\,2. 

We made the $^{12}$CO position-velocity (PV) diagram in the horizontal direction along the Main cloud to explore the velocity structure further (Figure~\ref{fig:mom1_pv}(b)). Although we see some velocity fluctuations, the Main cloud shows an almost single velocity component with the velocity width of $\sim$10--20\,km\,s$^{-1}$. We thus conclude that the Main cloud is almost composed of a single velocity component. However, at the western and eastern side of the Main cloud, there are notable blueshifted components whose velocity is $\sim-$260\,km\,s$^{-1}$, as appeared in Figure~\ref{fig:mom1_pv}(b). 
We confirmed large differences at most $\sim$15\,km\,s$^{-1}$ among the centroid velocities of $^{13}$CO clumps (Sect.~\ref{result:space}). 
The $^{13}$CO contours on the PV diagram in Figure~\ref{fig:mom1_pv}(b) demonstrate that clump-4 and -5 are at the edge of the $^{12}$CO emission rather than centrally concentrated along with the $^{12}$CO intense velocity range. We discuss these velocity features and the possibility that more than two velocity gas streams develop in the massive cloud in Sect.~\ref{dis:origin}.

\section{Discussion} \label{sec:dis}

\subsection{Comparisons with Other Massive GMCs in the Local Group}\label{dis:LoGMC}
As mentioned in the introduction, GMC-8 is one of the most massive GMCs in M33. The GMC belongs to a series of extremely massive clouds in the Local group of galaxies at least in terms of its total H$_2$ mass. This section summarizes previous studies of massive GMCs in the MW, LMC, and M33 from the literature, and then compares their physical properties and those of GMC-8. Table\,\ref{tab:GMC} shows the characteristics of massive GMCs based on the CO or its isotopologue observations with an angular resolution of 1--40\,pc and a velocity resolution of 0.1--0.6 \,km\,s$^{-1}$. The massive GMCs, whose molecular gas mass exceeds $\sim$10$^{6}$\,$M_{\odot}$ in the MW, actively form high-mass stars almost exclusively. The Maddalena's cloud and Cygnus~OB7 are rare clouds as quiescent GMCs, which are hard to find within $\sim$1\,kpc from the Sun. 

The sizes and velocity widths do not differ greatly among these GMCs except for a relatively narrow line width in Cygnus~OB7. We found a marginal indication of a higher line width in GMC-8. If the observational bias artificially produces such a feature, this effect should be more pronounced in the MW targets due to its edge-on view. 
Therefore, we speculate that GMC-8 inherently has a large line width, which means that a high-level turbulent motion dominates the cloud or there is a superposition of multiple velocity components. For example, spatially/velocity resolved observations of the MW/LMC/M33 massive clouds obtained substantial evidence of cloud--cloud collisions (W43, \citealt{Kohno21}; W49, \citealt{Miyawaki09}; W51, \citealt{Fujita21a}; Carina, \citealt{Fujita21b}; LMC-N159, \citealt{Fukui19,Tokuda19}; GMC-37, \citealt{Sano20}; NGC~604, Paper~II; GMC-16, Paper~I) and the authors suggest that such drastic events triggered high-mass star formation therein. Although \cite{Dobashi14} reported colliding filaments with a few km\,s$^{-1}$ velocity difference in Cygnus~OB7 at the dense cloud center, such local phenomena (i.e., filament collisions) does not largely change the GMC scale velocity width as a whole. 
Star-formation active GMC tends to show highly developed structures composed of many spatially steep features, such as shells and filaments, while an extended diffuse component dominates less active clouds (see also a quantitative analysis in the LMC; \citealt{Sawada18}). The extended gas feature of GMC-8 is qualitatively different from filamentary structures discovered in the other two GMCs in M33 (Papers~I, II). 
The low-density nature and less-developed structure are consistent with those in the other star-formation inactive clouds, but the velocity width is comparable to active star-forming regions in the Local group. We discuss the velocity structure and the formation origin of GMC-8 in Sect.~\ref{dis:origin}.

\begin{deluxetable*}{cccccc}
\tabletypesize{\scriptsize}
\tablewidth{0pt} 
%\tablenum{3}
\tablecaption{Physical Properties of GMCs in the Local Group
\label{tab:GMC}}
\tablehead{
\colhead{GMC Name} & \colhead{Hosting Galaxy} & \colhead{Size} & \colhead{Mass}${(1)}$  & \colhead{$v_{\rm FWHM}$}${(2)}$  & \colhead{References}\\[-1.0mm]
      & & (pc) &($M_{\odot}$)  & (km\,s$^{-1}$) & %\\[-1.0mm]
}
%\colnumbers
\startdata 
\hline
Active star-formation \\ \hline
W43        & Milky Way & $\sim$70$\times$150         & 2$\times$10$^{6}$ & 5--10    & \cite{Carlhoff13}  \\
W49        & Milky Way & $\sim$120$\times$120 & 1$\times$10$^{6}$ & 4--9     & \cite{Madrid13} \\
W51        & Milky Way & $\sim$80$\times$110  & 1$\times$10$^{6}$ & $\sim$9  & \cite{Carpenter98}  \\
Carina     & Milky Way & $\sim$60$\times$80   & 2$\times$10$^{5}$ & $\sim$5  & \cite{Rebolledo16}  \\
N159       & LMC       & $\sim$60$\times$100  & 6$\times$10$^{5}$ & 7--10    & \cite{Minamidani08} \\
NGC~604    & M33       & $\sim$100$\times$200 & 3$\times$10$^{6}$ & $\sim$9  & \cite{Muraoka20} \\
GMC-16     & M33       & $\sim$180$\times$80  & 2$\times$10$^{6}$ & $\sim$6  & \cite{Tokuda20} \\ \hline
Inactive star-formation\\ \hline
Maddalena's cloud  & Milky Way & $\sim$250$\times$100 & 1$\times$10$^{5}$ & $\sim$8  & \cite{Lee94}\\
Cygnus~OB7 & Milky Way & $\sim$80$\times$100  & 2$\times$10$^{5}$ & $\sim$4  & \cite{Dobashi94}\\
GMC225     & LMC       & $\sim$100$\times$40  & 1$\times$10$^{5}$ & $\sim$6  & \cite{Minamidani08} \\
GMC-8      & M33       & $\sim$70$\times$80   & 4$\times$10$^{6}$ & $\sim$11 & This work \\
\enddata
\tablecomments{(1) Total molecular mass traced by CO observations. (2) FWHM of the velocity profile.}
\end{deluxetable*}

\subsection{A possible formation origin and fate of GMC-8}\label{dis:origin}

Previous numerical simulations investigate molecular cloud formation and its mass supply by atomic gas (H$\;${\sc i}) supersonic flow \citep[e.g.,][]{Hennebelle99,Koyama00,Koyama02,Vazquez07,InoueInu12}. However, some theoretical works \citep[e.g.,][]{Kobayashi17,Kobayashi18} suggest that it takes $\sim$100\,Myr to gain the H$_2$ mass up to $\sim$10$^{6}$\,$M_{\odot}$ in the case of recursive accretion flows from multiple directions with a density of $\sim$1\,cm$^{-3}$ (see \citealt{Koda09}, for estimations considering a spherical configuration of the mass accretion, which also suggest a similar timescale of 100\,Myr). Such a long timescale is significantly larger than the typical GMC lifetime, a few $\times$ 10\,Myr \citep{Fukui99,Kawamura09}. \cite{Fukui09} observationally derived an H$\;${\sc i} gas accretion rate onto GMCs of $\sim$0.05\,$M_{\odot}$\,yr$^{-1}$ by considering the three-dimensional relation between the H$\;${\sc i} envelope and molecular clouds in the LMC. In this case, although it is possible to grow up to $\sim$10$^{6}$\,$M_{\odot}$ within the GMC lifetime, the star-formation inactive GMC is supposed to evolve into the subsequent state (i.e., inhering bright H$\;${\sc ii} regions) in a much shorter time, $\sim$6\,Myr \citep{Kawamura09,Corbelli17}. Therefore, in order to obtain a large amount of molecular gas in a shorter time, it is necessary to supply materials by a drastic phenomenon rather than a steady-state accretion condition.

\begin{figure}[htbp]
\begin{center}
\includegraphics[width=110mm]{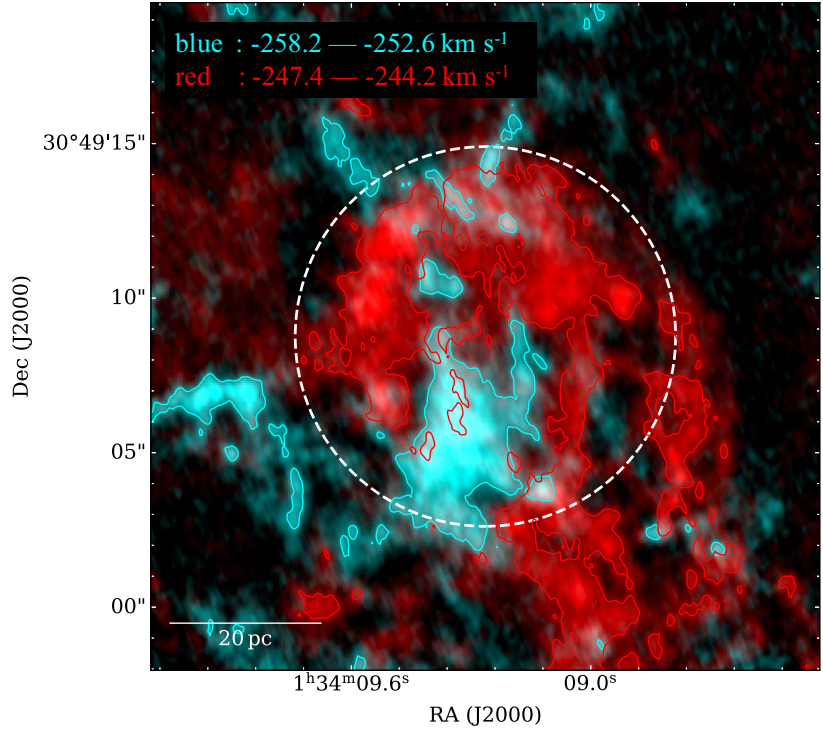}
\end{center}
\caption{Spatially complementary clouds with two different velocity ranges around the Main cloud. The red and cyan images show the integrated intensity of the $^{12}$CO ($J$\,=\,2--1) emissions whose velocity ranges are -247.4\,--\,-244.2\,km\,s$^{-1}$ and -258.2\,--\,-252.6\,km\,s$^{-1}$, respectively. The red and cyan contour levels are 8\,K\,km\,s$^{-1}$ ($\sim$8$\sigma$). The white dashed circle shows the arbitrarily defined region of the Main cloud (see the text). 
\label{fig:compl}}
\end{figure}

We revisit the PV diagram (Figure\,\ref{fig:mom1_pv}) to investigate whether or not the velocity structure preserves the remnants of a gas convergence. As we show in Sect.~\ref{result:veloicty}, the GMC is composed of an almost single velocity component with some local fluctuations. Among these features, the clump-4 is relatively blueshifted in the Main cloud (see panel (b)). This characteristic is similar to a synthetic PV diagram, which is called as $``$V-shaped$"$ feature, produced by cloud--cloud collision \citep{Takahira14,Haworth15,Fukui18}. Figure\,\ref{fig:compl} shows the blueshifted and redshifted CO components with complementary distributions with each other, which is also one of the pieces of evidence for gas collision \citep{Fukui21}. Based on these results, we set up a hypothesis that two molecular clouds with a relative velocity difference of $\sim$10\,km\,s$^{-1}$ (see the next paragraph) collided with each other, and the event realizes the total molecular mass of $\sim$10$^{6}$\,$M_{\odot}$ in this region. 

Notably, there are two independent blueshifted ($\sim-$260\,km\,s$^{-1}$) clouds at the east and west sides of the Main cloud (see Figure\,\ref{fig:mom1_pv} (b)). If the blueshifted colliding counterpart was larger than the GMC seed, the extended parts of blueshifted clouds could not interact with the Main cloud. We may witness the present blueshifted components as two remnant clouds that are not decelerated by the collision (see also Section \ref{dis:Hiflow}). The average column density of the Main cloud and blueshifted clouds are $\sim$3\,$\times$\,10$^{22}$\,cm$^{-2}$ and $\sim$1\,$\times$\,10$^{22}$\,cm$^{-2}$, respectively. If the column density ratio represents the mass ratio between the current product (i.e., the Main Cloud) and the GMC seed before the collision, the extra blueshifted component would contribute at least 20--30\% as the mass reservoir to produce the million-solar-mass GMC.

Even though it is not easy to precisely estimate the collision's timescale, the expected duration is on the order of a few Myr (= a few $\times$10\,pc/$\sim$10\,km\,s$^{-1}$), which is derived from the GMC size divided by the relative velocity. This is consistent with the statistically derived timescale of the star formation inactive phase without remarkable H$\;${\sc ii} regions. According to the compilation by \cite{Corbelli17}, the total number of such GMCs in M33 is 178 (= Type A + B clouds in their catalog). The fact that there are no other clouds with a comparable mass to GMC-8 at the same stage indicates that its lifetime is very short if we assume the ratio of the number of objects at each stage is proportional to that of the timescale. Possible fate of GMC-8 is to evolve into a less massive GMC immediately by (i) gas consumption/destruction through star formation and/or (ii) dissipating the molecular gas due to turbulence without star formation. The presence of the 24$\mu$m source at the vicinity of clump-5 indicates that an embedded star formation is already on-going. Meanwhile, as shown in Table\,\ref{tab:13COclump}, the virial mass of the $^{13}$CO clumps is larger than their $M_{\rm LTE}$ except for clump-5, indicating that most of the dense materials are not self-gravitating. Despite the presence of the collisional signatures in this system, the inactive nature of high-mass star-formation can be explained by the lower column density of the molecular gas that can produce at least one single O-star. According to the recent statistical investigations of colliding clouds among more than 50 samples by \cite{Enokiya21}, at least a single O-star formation requests higher H$_2$ column density more $\sim$10$^{22}$\,cm$^{-2}$, which is mostly higher than those of the $^{13}$CO clumps in GMC-8. Recent ALMA observations toward the interacting galaxies system, NGC~4567/4568, also found a star-formation inactive molecular layer with a colliding velocity feature \citep{Kaneko18}. In summary, the combination of the local star-formation activity and gas dissipation due to turbulent motion may be going to reduce the molecular gas of this system in the near future. 

\subsection{Comparison with Large-scale H$\;${\sc i} and CO Distributions}\label{dis:Hiflow}

We subsequently discuss the origin of the colliding gas whose relative velocity is $\sim$10\,km\,s$^{-1}$ against the Main cloud. Supernova remnant searches in M33 \citep[e.g.,][]{Garofali17,Long18} did not find any candidates within $\sim$300\,pc around GMC-8, which means that supernova feedback is unlikely to be the origin of the high-velocity colliding flow. The location of the GMC is in the interarm region (see Section~\ref{sec:intro}), and thus the galactic shock also cannot be appropriate, as suggested in our GMC-16 study (Paper~I) whose location is near the stellar spiral arm. 

\begin{figure}[htbp]
\includegraphics[width=180mm]{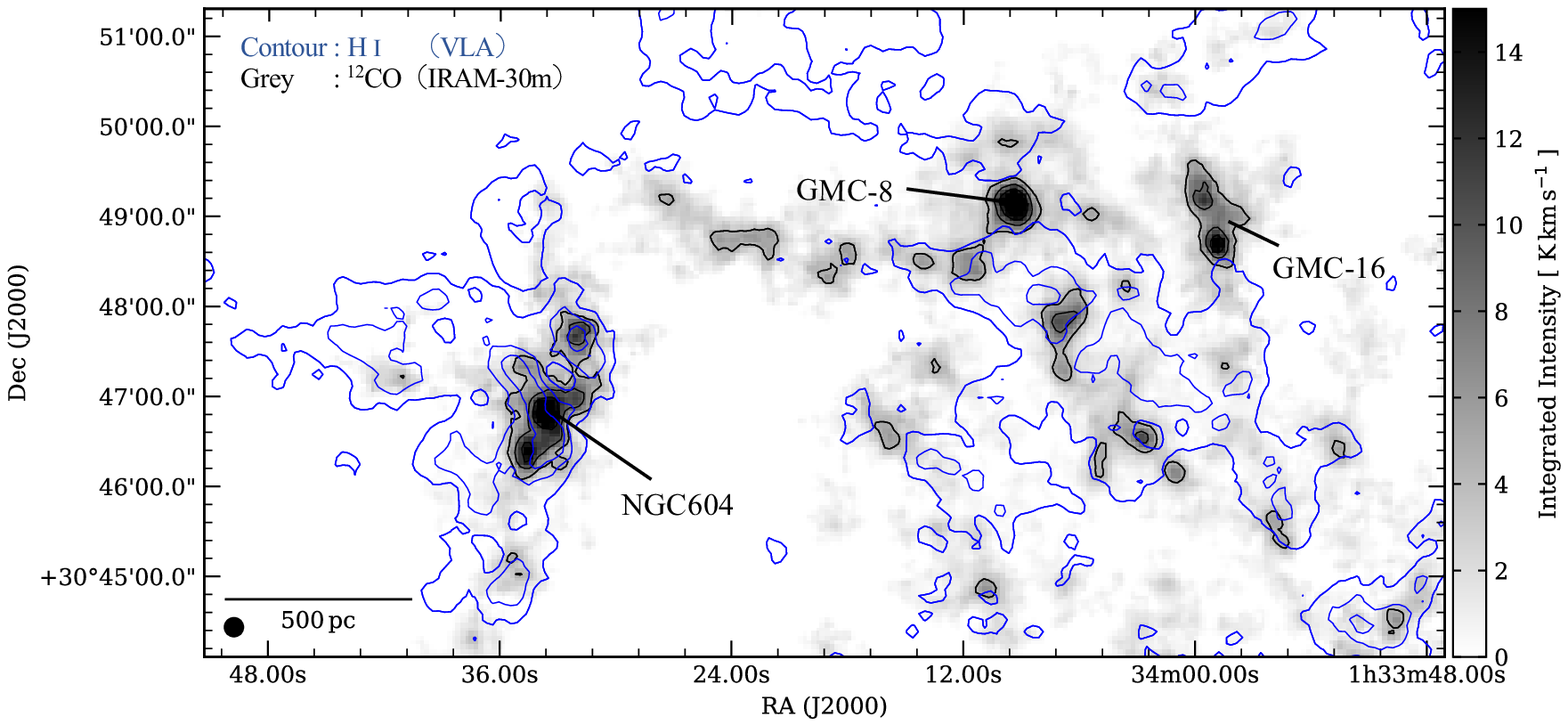}
\caption{The gray-scale image and black contour show the integrated intensity of $^{12}$CO($J$\,=\,2--1) with the IRAM 30\,m telescope \citep{Druard14}. The lowest contour and subsequent levels are 2\,K\,km\,s$^{-1}$. The blue contours show the blue-shifted H$\;${\sc i} clouds with the integrated ranges of $\sim-$214\,km\,s$^{-1}$\,\textless\,$V_{\rm shift}$\,\textless\,$\sim-$184\,km\,s$^{-1}$ (see \citealt{Tachihara18}). The lowest and subsequent contour steps are 200\,K\,km\,s$^{-1}$, and 300\,K\,km\,s$^{-1}$, respectively. The black circle at the lower left corner represents the angular resolution of the CO image, $\sim$11\arcsec.
\label{fig:HI_CO}}
\end{figure}

\begin{figure}[htbp]
\begin{center}
\includegraphics[width=150mm]{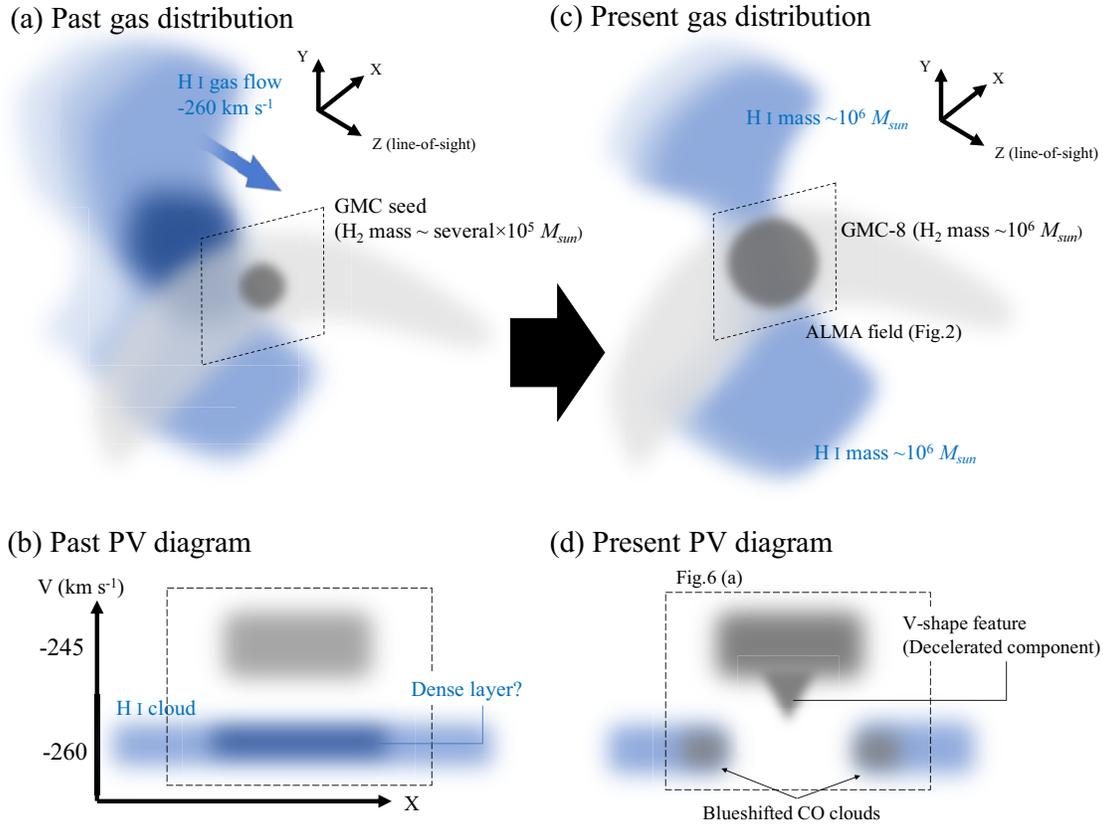}
\end{center}
\caption{Schematic view of past/present gas distributions and cartoon position-velocity diagrams of GMC-8.
\label{fig:ponti}}
\end{figure}

We here investigate a large-scale H$\;${\sc i} gas distribution obtained with the Very Large Array (VLA) observations \citep{Gratier10}. We used a velocity cube that we corrected to the the galactic rotation, and the shifted velocity frame is $V_{\rm shift}$ (see \citealt{Tachihara18}). We overlay a blueshifted velocity H$\;${\sc i} component ($V_{\rm shift}$ = $-$214--$-$184\,km\,s$^{-1}$), which is the same as the NGC~604 study defined by \cite{Tachihara18}, on the CO gas (Figure\,\ref{fig:HI_CO}). This velocity is consistent with that in the blueshifted ($-$260\,km\,s$^{-1}$) CO gas (Figure~\ref{fig:mom1_pv}) in the standard frame.
The colliding H$\;${\sc i} gas is likely converted into a molecular state at the GMC position as a whole. Our approximate estimation of the blueshifted atomic gas mass at both of the northern/southern sides of GMC-8 is $\sim$10$^{6}$\,$M_{\odot}$ within the same area as GMC-8 traced by the single-dish CO observation assuming the optically thin limit of the H$\;${\sc i} emission. If we assume such a large amount of the atomic gas as an initial condition of the colliding H$\;${\sc i} flow onto a GMC-8 seed, a certain fraction of hydrogen gas may be converted into H$_2$ during the collision. 
Figure~\ref{fig:ponti} summarizes our suggestion for the mass growth process of GMC-8. There were a less massive GMC seed along the interarm region and a blueshifted H$\;${\sc i} gas component possibly containing a dense layer before the collision (panels (a) and (b)). Panels (c) and (d) mimic observed gas distribution in $X-Y-Z$ and the PV diagram. The deceleration by the collision of the two clouds explains the lack of blueshifted components toward the Main cloud (see Sect.\,\ref{dis:origin}). The present circumstance around GMC-8 suggests that the large-scale ($\gtrsim$50\,pc) H$\;${\sc i} flow is the promising phenomena to increase GMC mass up to $\sim$10$^{6}$\,$M_{\odot}$. This is consistent with the theoretical prediction that a dynamic mass accumulation is required to form such massive GMC rather than local converging flows (see the introduction).

We finally remark on the origin of the H$\;${\sc i} flow. \cite{Tachihara18} pointed out that a tidal interaction between M31 and M33 drove external H$\;${\sc i} flows onto the M33 disk based on the presence of a bridge feature connecting the two galaxies \citep{Braun04,Putman09,Lockman12}. \cite{Grossi08} found high-velocity H$\;${\sc i} clouds, which are supposed to be tidal debris, and suggested they are supply sources for high-mass star formation. Alternatively, we cannot exclude an internal origin of such high-velocity diffuse gas. \cite{Wada11} proposed that the dynamic spiral theory, which involves nonsteady stellar arms (\citealt{Dobbs14} and references therein), explain the presence of complex gas structures in flocculent spiral galaxies like M33. In this case, we predict that there are gas flows from multiple directions onto molecular clouds. There are some high-velocity components apart from the galactic rotation other than GMC-8 and NGC~604 (Figure~\ref{fig:HI_CO}). Though we cannot determine which external or internal origins dominate as such gas supply phenomena, further investigating the relation between H$\;${\sc i} and CO will give a better understanding of GMC evolution during the galactic lifecycle.

\section{Summary} \label{sec:sum}
This paper describes ALMA observations of one of the star-formation inactive GMCs in M33, GMC-8, with a total H$_2$ mass of $\sim$10$^{6}$\,$M_{\odot}$. As one of the most massive GMCs in the galaxy, it is a vital target to understand such an exclusive target's formation mechanism in the Local group of galaxies. The primary conclusions are summarized as follows:

\begin{enumerate}
\item The molecular gas (traced by $^{12}$CO) distribution in an $\sim$1\,pc view is a highly extended and round-shaped structure. This finding is different from our companion studies (Papers~I, II) of two active star-forming GMCs with many remarkable shell/filamentary structures. 

\item The $^{13}$CO total flux and peak intensity of GMC-8 are remarkably weaker than those in the other two GMCs. The dense ($\sim$10$^{3}$--10$^{4}$\,cm$^{-3}$) gas fraction with respect to the total H$_2$ mass judging from the $^{13}$CO/$^{12}$CO data is as low as 2\%. The development of density and structure in molecular clouds is necessary for the process leading to high-mass star formation. 

\item Our compilation from the literature found that the velocity width of GMC-8 is similar to or slightly higher than those of other active high-mass star-forming GMCs in the Local universe. Although the GMC velocity component appears to be single as a whole, the velocity/spatially resolved observations found several characteristics that coalescence of two components forms the molecular cloud. We suggest that a convergent gas flow onto the GMC seed with a molecular mass of several $\times$10$^{5}$\,$M_{\odot}$ may work to grow up at most $\sim$10$^{6}$\,$M_{\odot}$ without any remarkable high-mass star formation activities on a relatively short timescale of a few megayears.
\end{enumerate}

\acknowledgments
This paper makes use of the following ALMA data: ADS/ JAO.ALMA\#2017.1.00461.S. ALMA is a partnership of the ESO, NSF, NINS, NRC, NSC, and ASIAA. The Joint ALMA Observatory is operated by the ESO, AUI/NRAO, and NAOJ. This work was supported by NAOJ ALMA Scientific Research grant No. 2016-03B and JSPS KAKENHI (grant Nos. JP17K14251, JP18K13580, JP18K13582, JP18K13587, JP18H05440 and JP20H01945). Dr. Pierre Gratier kindly provided us with the VLA H$\;${\sc i} data cube \citep{Gratier10} to discuss the relation between the large-scale diffuse gas and the CO GMC. T. T is AOJ ALMA Scientific Research grant 2020-15A.
\software{CASA (v5.4.0; \citealt{McMullin07}), Astropy \citep{Astropy18}, APLpy v1.1.1 ;\citep{Robi12}}

%\restartappendixnumbering
\appendix

%\restartappendixnumbering

\bibliography{sample63}{}
\bibliographystyle{aasjournal}

\end{document}